# THE NATURE VERSUS NURTURE
# OF ANISOTROPIES

WAYNE HU

*Departments of Astronomy and Physics, University of California*
*Berkeley, California 94720, USA*

ABSTRACT

With the rapidly growing number of cosmic microwave background anisotropy measurements on various scales, there is real hope that the number of acceptable models for structure formation will be limited to a very few in the near future. Yet any given model can always be saved by introducing and tuning extraneous free parameters. To better understand this question of "nature versus nurture" for temperature fluctuations, it is useful to know not only the general features of anisotropy predictions but also their causes. Extracting the physical content of our other works, we present here a *simple* account of anisotropies on all scales. In particular, we show that analytic approximations can trace the structure of the so-called "Doppler peaks," which arise due to the *adiabatic* oscillations in the photon-baryon fluid. We also show how the finite thickness of the last scattering surface and the Silk damping mechanism can be described in a unified way by photon diffusion. In order to present a specific example, we focus on comparing the primordial isocurvature baryon (PIB) model with the standard cold dark matter model (CDM). In particular, we explain why PIB generically predicts larger *non*-oscillatory anisotropies from the 1° to 10° scale which may already be in conflict with experiments.

## 1. Introduction

*What comes from heaven is caused to be unique. It is man who disguises and assimilates. By this I know it is from heaven not man...*

                                                        –*Chuang-tzu*

The standard cold dark matter (CDM) model is impressive both for its simplicity and for its ability to predict roughly the right phenomena over the vast range of scales from the present horizon down to galaxies. Can we view this as evidence that some variant of the standard CDM model is correct? If the detection of the so-called "Doppler peaks" in the degree scale cosmic microwave background (CMB) is confirmed,[1] should it be viewed as convincing support for the general picture? In short, can significantly different models mimic the desirable features of the CDM model?

Of course, the answer to the last question is "yes," if we are willing to feed in as many free and arbitrary parameters as there are observational constraints! The question is then partially one of aesthetics. To answer it, we need to know not only what sort of anisotropies are natural for a given model but also *why*. Although the general behavior of anisotropies is well known for most models, obtaining such results conventionally involves the use of a black box: the cosmological Boltzmann code.[2,3] With the purpose of clarifying the nature of anisotropies in mind, we have here distilled the physical essence of our work on a purely *analytic* understanding of anisotropies.[4,5] Whereas we have elsewhere developed a detailed treatment which obtains the anisotropies on all scales to 10-15% in temperature fluctuations, here we aim only for a *simple* explanation of the broad features of the anisotropies. The discussion contained here should be viewed as a qualitative summary and preview of our other works.

It is useful to have a specific example to contrast with the standard CDM model. In 1987, Peebles[6] proposed just such an alternative: the primordial isocurvature baryon (PIB) model. This model not only satisfies the general requirements for structure formation without requiring hypothetical exotic dark matter, but also respects dynamical observations of a low density $\Omega_0 < 1$ universe. This "good thing" is to be balanced with the undesirable prediction of a large baryon fraction $\Omega_b = \Omega_0 \approx 0.1 - 0.3$ which requires significant alteration of big bang nucleosynthesis.[7] Unfortunately though, this original form of the PIB scenario has also been recently ruled out[8] by the low upper limit on spectral distortions in the CMB by COBE FIRAS.[9] Can the PIB scenario be resurrected without introducing unnatural and *ad hoc* features? If so, can its new incarnation be distinguished from the standard CDM model?

In these *Proceedings*, we will provide an informal discussion of these issues and focus on qualitative results. More quantitative analyses can be found in our detailed treatments.[4,5] In §2, we examine the underlying unity behind mechanisms of anisotropy generation for both models, while highlighting their differences. We find that simple arguments as to the nature of the Sachs-Wolfe, adiabatic, and Doppler effects as well as photon diffusion damping can account for the structure of anisotropies in all gravitational instability models including even their dependence on cosmological parameters. In §3, we discuss to what extent predictions of the PIB model can and cannot be altered to suit observations by tuning the thermal history and initial conditions. Larger fluctuations between the COBE DMR and degree scales than the CDM scenario is a relatively robust prediction in the PIB model, thus serving to make the theory verifiable. Indeed, its fate rests in the hands of *present day* experiments.

## 2. The Nature of Anisotropies

It is no surprise that these two models, PIB with baryons only and isocurvature initial conditions and CDM with cold non-interacting matter and adiabatic initial conditions have similar causes for temperature anisotropies. After all, both form structure through gravitational instability. Until recent epochs, where possible curvature of the universe in the PIB scenario affects the dynamics, the only difference between the two is the initial conditions. In open CDM models, even this distinction disappears.

It is evident that the *evolution* of fluctuations for both the matter and the CMB obeys the same principles. In particular, temperature anisotropies in both models can be broken up into three parts: (1) the Sachs-Wolfe effect at large scales, (2) the adiabatic effect at the intermediate scale, (3) the Doppler effect at small scales. Crudely speaking, intermediate scales can be considered as those falling between the horizon at last scattering and the thickness of the last scattering surface. In the standard recombination scenario, where photons last scattered at $z_* \approx 1000$, degree scale anisotropies correspond to the *adiabatic* effect. Tight coupling between the photons and baryon makes the thickness scale, i.e. the photon diffusion scale at last scattering, much smaller than the horizon, leaving a large window for the adiabatic effect. On the other hand reionized scenarios, where last scattering is comparatively recent and the coupling correspondingly weakened, can be dominated by the Doppler effect at degree scales. We shall see that reionization is far more likely in the PIB than the CDM scenario.

### 2.1. Initial Conditions

For any set of initial conditions, the early evolution of perturbations is marked by *adiabatic* growth and decay of the photon and baryon energy density fluctuations. Specifically, the entropy perturbation, defined by spatial fluctuations in the baryon to photon number density ratio,

$$\delta(n_b/n_\gamma) = \Delta_b - {}^3\!/_4 \Delta_\gamma \equiv S, \qquad (1)$$

is constant, $\dot{\Delta}_b - {}^3\!/_4 \dot{\Delta}_\gamma = \dot{S} = 0$, where $\Delta_b$ and $\Delta_\gamma$ are the (gauge invariant) energy density perturbations. Since the photons and baryons are tightly coupled by Compton scattering before last scattering, these particles cannot stream away from each other. Consequently, their respective energy densities must evolve together. Even for cold dark matter or massless neutrinos, causality prevents the components from separating on superhorizon scales, and the evolution is still adiabatic. We will use the terms "tight coupling" and "adiabatic evolution" interchangeably.

Whereas in the adiabatic scenario the constant $S = 0$, in an isocurvature model the universe begins with a spatial fluctuation in the equation of state which gives the desired entropy fluctuation. Since as yet no mechanism exists for generating isocurvature fluctuations of the right spectral slope for structure formation, there is no rigorous way to fix the power spectrum of the initial conditions. For simplicity, it is often taken to be a pure power law in scale[10] $\tilde{k}$, $S^2(\tilde{k}) \propto \tilde{k}^n$ where $\tilde{k}^2 = k^2 + K$ and the spectral index is a free parameter adjusted to best fit the large scale structure data ($n \approx -1$).[11,12]

As an aside, we should briefly note that the initial conditions question raises a potentially important caveat for calculations of the PIB model since the universe is open. As is well known, when one decomposes perturbations in the open universe equivalent of plane waves (i.e. eigenfunctions of the Laplacian) of eigenumber $k$, *no* power spectrum of these modes can produce significant structure above the curvature scale $\lambda > 1/\sqrt{-K}$ where the curvature $K = -H_0^2(1 - \Omega_0)$. We would like here to set aside formal mathematical questions as to the nature of spectral decomposition in non-Euclidean geometries and show how this oddity of open universes can be phrased in purely physical terms. To do this, let us suppose that the perturbations have support only over a finite region which nonetheless extends *much larger* than the curvature scale (and the present horizon). Even here structure larger than the curvature scale is exponentially suppressed, if we assume random fields defined by *any* power spectrum. Yet since the eigenfunctions of the Laplacian are complete, larger than curvature scale correlations *can* be expressed in terms of these functions. The point is that no random superposition of them will ever produce structure larger than the curvature scale. An examination of the functional form of the radial eigenfunctions clarifies this oddity: they all die exponentially above the curvature scale. Only a conspiracy of modes can ever produce structure above the curvature scale that is not likewise exponentially suppressed. This is closely related to the fact that the volume grows exponentially above the curvature scale. As noted by Kamionkowski & Spergel,[13] if fluctuations have a power law scaling in volume, structure above the curvature scale *should* be exponentially suppressed.

The question is therefore a purely physical one: are fluctuations generated with random phases for the eigenmodes of the Laplacian? The answer is "yes" for fluctuations generated by inflation[14,15] since the eigenmodes of the Laplacian form a natural basis. While we have every reason to expect the same is true for isocurvature fluctuations, our present lack of knowledge as to the mechanism of their generation makes a definitive statement impossible. The work on open inflationary universes also drives home another cautionary point: the power spectrum may not be a single power law, i.e. the presence of curvature inserts a natural scale into the problem. We will argue in §3.2 that this is not a severe uncertainty in the predictions for PIB, but it is important to keep this caveat in mind.

*2.2. Early Evolution*

Questions as to the nature of the initial spectrum and mode correlations discussed above do not affect the evolution of perturbations, because each mode evolves independently in linear theory. Since superhorizon scale modes in the adiabatic scenario evolve in a simple and well known manner, we will concentrate here on the isocurvature case. As the name implies, isocurvature fluctuations are generated with vanishing perturbations to the curvature, i.e. the total density perturbation $\Delta_T = \delta\rho_T/\rho_T \approx 0$ initially. Since the universe is radiation dominated early on, this is arranged by placing the fluctuations in the baryons with little compensation from the photons $\Delta_b = S$, $\Delta_\gamma \approx 0$. This balance, once set up, is partially preserved through the evolution into the matter dominated epoch, but as we shall see it is ultimately frustrated by the need to maintain adiabatic evolution.

As the universe evolves, the relative significance of the baryon compared with the radiation fluctuation grows as $a/a_{eq}$ where equality occurs at $a_{eq} = 4.0 \times 10^{-5}(T_0/2.7\text{K})^4(\Omega_0 h^2)^{-1}$ with $T_0 = 2.726 \pm 0.010$K as the present temperature of the CMB,[9] and a Hubble constant of $H_0 = 100h$ km s$^{-1}$ Mpc$^{-1}$. To compensate, the photon fluctuations must grow to be equal and opposite $\Delta_\gamma = -(a/a_{eq})S$. Tight coupling implies then that the baryon fluctuation must also decrease so that $\Delta_b = (1 - 3a/4a_{eq})S$. The presence of $\Delta_\gamma$ means that there is a gradient in the photon energy density. This gradient gives rise to a dipole $V_\gamma$ as the regions come into causal contact, i.e. $V_\gamma \propto -k\eta\Delta_\gamma \propto -k(a/a_{eq})^2 S$ where the conformal time, $\eta = \int dt/a \propto a/a_{eq}$ in the radiation dominated limit, gives the horizon size. Tight coupling requires that the total fluid move with the photons $V_T = V_\gamma$, and thus infall, produced by the gradient in the velocity, yields a total density perturbation $\Delta_T \propto -k\eta(1 - 3K/k^2)V_T \propto (k^2 - 3K)(a/a_{eq})^3 S$. The curvature correction to the total matter continuity equation used here can be constructed from the definition of the $\Delta$'s and accounts for the curvature term in the Hubble parameter.[5]

This is the physical explanation for the well known result that the entropy $S$ provides a source of total density fluctuations which evolve as $(k^2 - 3K)(a/a_{eq})^3 S$ in the radiation dominated limit.[16] A similar analysis applies for adiabatic fluctuations, which begin with total fluctuations $\Delta_T \propto \tilde{k}^{n/2}$ and a corresponding gravitational potential $\Psi \propto \tilde{k}^{n/2}(k^2 - 3K)^{-1}$. Infall implies $V_T \propto k\eta\Psi$ which then yields $\Delta_T \propto k\eta V_T \propto k^{n/2}(a/a_{eq})^2$. Notice that compared to the adiabatic case, the isocurvature scenario predicts total density perturbations which are smaller by one factor of $a/a_{eq}$ as might be expected from cancellation. Furthermore, $\Delta_T$ becomes dominant over the entropy $S$ when the perturbation enters the horizon in the matter dominated epoch. In Fig. 1, we depict this evolutionary mechanism for an $\Omega_0 = 0.2$ and $h = 0.5$ model and compare analytic results[4] with the full numerical solution.[3] Behavior after horizon crossing at $a_H$ is discussed in §2.4 and after last scattering at $a_*$ in §2.3.

Regardless of initial conditions, when the universe becomes *fully* matter dominated the total density perturbations evolve as a pressureless fluid, i.e. $\Delta_T \propto a$ in the matter dominated epoch and remains constant in the curvature dominated epoch $z \lesssim z_g = (1/\Omega_0 - 1)$. It does not matter whether the fluctuations were generated in the radiation dominated epoch by the isocurvature mechanism or the usual adiabatic one. However, it is important to note that the isocurvature mode has gained a $k^2 - 3K$ tail from evolution in the radiation dominated evolution. Thus far we have neglected the effect of photon pressure which will prevent growth of fluctuations below the Jeans scale. We will discuss this further in §2.4 where we treat adiabatic oscillations. We shall see that those considerations and the $k^2 - 3K$ tail found here imply that there will be a prominent peak in the matter transfer function for PIB at the maximal Jeans scale.

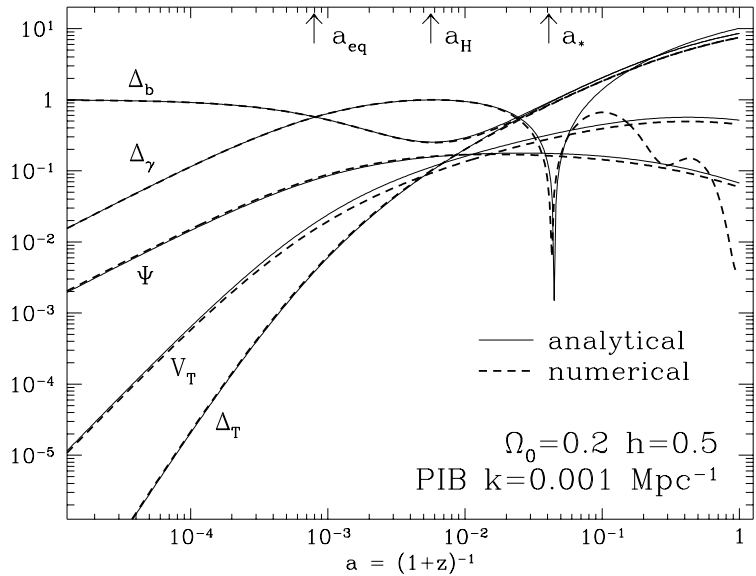

**Figure 1.** Early isocurvature perturbation evolution. The universe begins with fluctuations in the baryons which are gradually transferred to the photons as the universe becomes matter dominated. A total density (curvature) fluctuation is a residual effect of this process. It comes to dominate over the entropy perturbation by horizon crossing at $a_H$ in the matter dominated epoch. After last scattering at $a_*$, photon energy density perturbations are transferred to anisotropies, whereas the matter fluctuations grow in linear theory.

This simple analysis brings out interesting consequences for the Newtonian potential defined through the generalized Poisson equation

$$\Psi = -\frac{4\pi G \rho}{k^2 - 3K} a^2 \Delta_T,$$

where we have assumed that anisotropic pressure is negligible. In reality, its inclusion, through the quadrupole moment of the neutrinos, would make a 10% correction to the temperature anisotropies. We treat such subtleties elsewhere.[4,5] For adiabatic initial conditions, outside the horizon $\Psi$ is a (different!) constant in the radiation and matter dominated limits, whereas for isocurvature conditions it grows as $a/a_{eq}$ in the radiation dominated limit to reach a constant in the matter dominated epoch.

It is worthwhile to note that although for adiabatic initial conditions, $\Delta_T^2(\eta = 0) \propto \tilde{k}^n$ implies $\Psi^2 \propto \tilde{k}^{n-4}(1-4K/\tilde{k}^2)^{-2}$, for isocurvature conditions the extra factor of $\tilde{k}^2 - 4K$ in $\Delta_T$ makes $\Psi^2 \propto \tilde{k}^n$. This brings out two important points regarding large scale anisotropies since they are generated by $\Psi$ (see §2.3). The isocurvature spectral index equivalent to an adiabatic index $n$ is actually $n + 4$. Secondly, in an open universe, adiabatic potentials suffer a cutoff above the curvature scale $\tilde{k} < \sqrt{-K}$ which is *not* present in the isocurvature spectrum. This also has interesting consequences for the shape of large scale CMB anisotropies[5] (see §3.2). In both cases, if the universe is open, the potential would decay after curvature domination, since $\Delta_T$ ceases to grow, leading to additional contributions to the anisotropy.

*2.3. The Sachs-Wolfe Effect*

In both the adiabatic and isocurvature scenario, large scale CMB anisotropies can be *completely* described by the Sachs-Wolfe effect which is in turn governed by the Newtonian potential $\Psi$. In particular, we shall see how the coefficient $1/3$ in the adiabatic effect and 2 in the isocurvature effect arises naturally as a combination of initial conditions and the integrated part of the Sachs-Wolfe (ISW) effect. For the purposes of the Sachs-Wolfe effect, it is convenient to introduce the (gauge invariant) variable $\Theta$ which corresponds to temperature perturbation $\Delta T/T$ for Newtonian hypersurface slicing. On superhorizon scales, we can ignore the effect of coupling to the baryons and describe the evolution of $\Theta$ as

$$\frac{d}{d\eta}[\Theta + \Psi](\eta, \mathbf{x}, \boldsymbol{\gamma}) = 2\dot{\Psi}, \qquad (2)$$

where $\gamma_i$ are the direction cosines for the photon momenta, and we have again dropped the correction due to anisotropic stress. The right hand side of equation (2) is the ISW effect[17] and accounts for the gravitational redshift in a time dependent potential. As we have seen, in the matter dominated epoch (before curvature domination) $\Psi$ is a constant in both models. Therefore, the quantity $\Theta + \Psi$ is conserved along a geodesic $\mathbf{x}(\eta)$ and in the absence of scattering after $\eta_*$, leads to present day anisotropies,

$$\Theta(\eta_0, \mathbf{x}_0, \boldsymbol{\gamma}_0) = \Theta(\eta_*, \mathbf{x}_*, \boldsymbol{\gamma}_*) + [\Psi(\eta_*, \mathbf{x}_*) - \Psi(\eta_0, \mathbf{x}_0)], \qquad (3)$$

if last scattering occurs in the matter dominated epoch, *and* we are still in the matter dominated epoch today at $\eta_0$. This has the simple interpretation that the temperature fluctuation today is its value at the last scattering epoch $\eta_*$ modified by the gravitational redshift due to the difference in potential between last scattering and the present. Since the potential at the present just contributes to the unobservable monopole fluctuation, the relevant quantity is $[\Theta + \Psi](\eta_*, \mathbf{x}_*, \boldsymbol{\gamma}_*)$. This of course is the well-known ordinary Sachs-Wolfe effect.[17]

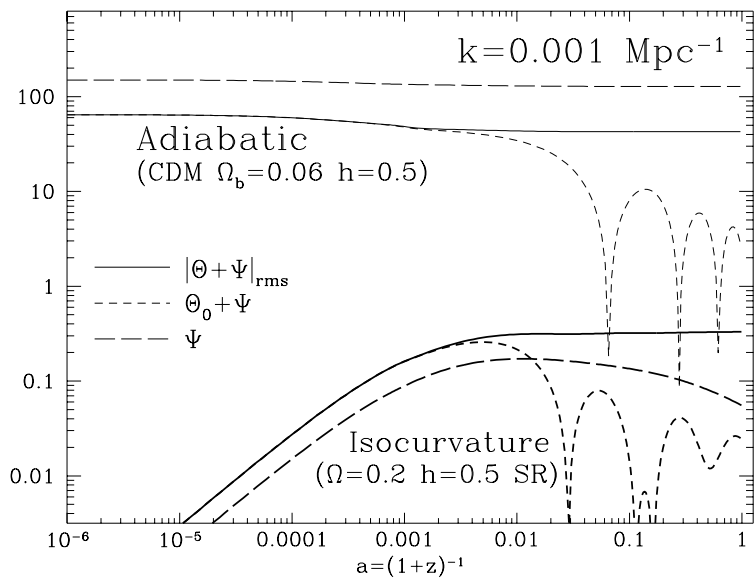

**Figure 2.** The CDM and PIB Sachs-Wolfe effect. Adiabatic initial conditions imply a constant potential in the matter and radiation dominated regimes with only a slight change at equality. This small correction through the ISW effect brings the final Sachs-Wolfe perturbation $|\Theta_0 + \Psi|$ to its well known value of $1/3\Psi$. The isocurvature Sachs-Wolfe effect is generated solely from the ISW effect and the growth of the potential. Notice that whereas the temperature and potential are anticorrelated in the adiabatic case, they are correlated in the isocurvature case. After last scattering monopole fluctuations are converted to anisotropies, leaving the rms temperature fluctuations constant until curvature domination.

We must now determine $\Theta + \Psi$ at last scattering. Equation (2) tells us,

$$\Theta(\eta_*, \mathbf{x}, \boldsymbol{\gamma}) \approx \Theta_0(0, \mathbf{x}) + \int_0^{\eta_*} \dot{\Psi}(\eta, \mathbf{x}) d\eta = \Theta_0(0, \mathbf{x}) + \Psi(\eta_*, \mathbf{x}) - \Psi(0, \mathbf{x}), \qquad (4)$$

if the perturbations are superhorizon scaled so that photons travel negligibly $\mathbf{x}(\eta_*) \approx \mathbf{x}(0)$. Since streaming or interactions have yet to produce an anisotropy, we have replaced $\Theta(\eta, \mathbf{x}, \boldsymbol{\gamma})$ with $\Theta_0(\eta, \mathbf{x})$, the isotropic (but inhomogeneous) monopole component. As an aside, notice that the sign of the potential difference in equation (4) is opposite to the ordinary Sachs-Wolfe effect. The ISW contribution, intrinsically a time dilation effect, cannot be reduced in a simple-minded way to an energy conservation argument.

Thus the ordinary Sachs-Wolfe effect breaks down into an initial conditions problem and the ISW effect. For isocurvature initial conditions both the temperature and potential fluctuations are negligible early on, and the ordinary Sachs-Wolfe effect becomes

$$[\Theta + \Psi](\eta_*, \mathbf{x}, \boldsymbol{\gamma}) = 2\Psi(\eta_*, \mathbf{x}) \qquad \text{ISO.}$$

Notice that $\Theta$ and $\Psi$ are of the same sign. This represents the fact that the isocurvature mechanism described in §2.2 *anticorrelates* $\Delta_\gamma$ and $\Delta_T$. For adiabatic initial conditions, the temperature perturbation starts out with the opposite sign from the potential $\Theta_0(0,\mathbf{x}) = -1/2\Psi(0,\mathbf{x})$, due to the fact that densities are higher deeper in the potential well. The potential does in fact change slightly through matter radiation equality so that $\Psi(\eta_*,\mathbf{x}) = 9/10\Psi(0,\mathbf{x})$. Together this implies

$$[\Theta + \Psi](\eta_*,\mathbf{x},\boldsymbol{\gamma}) = \frac{1}{3}\Psi(\eta_*,\mathbf{x}_*) \qquad \text{ADI}.$$

In Fig. 2, the Sachs-Wolfe contributions for adiabatic and isocurvature initial conditions are compared.

If the universe is not sufficiently matter dominated at last scattering as in the case of low $\Omega_0$ or $h$ models *or* is curvature dominated at the present, the integrated Sachs-Wolfe effect again arises. Both of these complications arise for the PIB model. Equation (3) must therefore be modified to read

$$\Theta(\eta_0,\mathbf{x}_0,\boldsymbol{\gamma}_0) = \Theta(\eta_*,\mathbf{x}_*,\boldsymbol{\gamma}_*) + [\Psi(\eta_*,\mathbf{x}_*) - \Psi(\eta_0,\mathbf{x}_0)] + \int_{\eta_*}^{\eta_0} 2\dot{\Psi}(\eta,\mathbf{x}_\eta)d\eta.$$

The integral over the photon geodesic $\mathbf{x}_\eta = \mathbf{x}(\eta)$ can be readily handled by decomposing perturbations in the radial eigenfunctions of the Laplacian (i.e. spherical Bessel functions for a flat universe). Detailed calculations can be found in several recent papers[18,19] as well as many earlier works.[3,20]

*2.4. The Adiabatic Effect*

When the photon-baryon perturbation enters the Jeans length where pressure dominates over gravity, photon pressure converts them into adiabatic oscillations of the photon-baryon fluid. This is the case for *both* adiabatic and isocurvature initial conditions and indeed for *any* mechanism of structure formation in which the baryon-photon fluid is perturbed before recombination. Thus the adiabatic effect exists in all model of structure formation that involve the gravitational instability of seed perturbations present before last scattering.

As has been pointed out previously,[21,22] the so-called "Doppler peaks" of the standard recombination scenario are somewhat misnamed since they are in fact due to adiabatic oscillations of the monopole $\Theta_0$. We would like to suggest that the whole structure of peaks and troughs should all be conceptually considered an adiabatic effect since they arise from adiabatic oscillations. To reiterate, the designation "adiabatic" refers to the fact that the entropy fluctuation $S$ remains constant, i.e. since the velocities of the two fluids are equal, they do not separate. The name "Doppler effect" is misleading even when we are speaking of the CMB dipole or "velocity." Since the photon and baryon velocities are equal, last scattering has no effect on the photons. It merely freezes in the oscillations. Until we get to the photon diffusion scale, where tight coupling begins to break down, there is no special imprint of the electron velocities on the photons *at* last scattering. In standard recombination scenarios, where the diffusion length is well under the scale that causes degree sized anisotropies in the CMB, we can ignore the true Doppler effect. We will come back to it when we consider reionized scenarios.

This tight coupling assumption allows us to describe the evolution of temperature perturbations with a *single* second order differential equation for the monopole,

$$\ddot{\Theta}_0 + \frac{\dot{a}}{a}\frac{R}{1+R}\dot{\Theta}_0 + \frac{1}{1+R}\frac{k^2}{3}\Theta_0 = F(\eta), \tag{5}$$

where overdots are derivatives with respect to conformal time, $R = 3\rho_b/4\rho_\gamma$, and the forcing function arises from the gravitational potential

$$F(\eta) = \ddot{\Psi} + \frac{\dot{a}}{a}\frac{R}{1+R}\dot{\Psi} - \frac{k^2}{3}\Psi. \tag{6}$$

We have extended the traditional tight coupling analysis[21,23] by including a realistic *time dependent* potential. This is necessary since the first few Doppler peaks occur right at the Jeans scale where gravity cannot be neglected. Notice also that in the limit that the scale is superhorizon-sized $k\eta \ll 1$, this equation merely returns the (integrated) Sachs-Wolfe effect $\dot{\Theta}_0 = \dot{\Psi}$. It is therefore applicable to the full range of scales until the breakdown of tight coupling at the diffusion damping scale (see §2.6).

Here we consider the gravitational term $F(\eta)$ to be a known *external* potential. For CDM, the potential is given by the decoupled dark matter in the matter dominated epoch; for PIB it is asymptotically given by the entropy perturbation $S$ as we shall see. Bear in mind that the potential on these scales can change significantly from their initial values even in the CDM model since $\Delta_T$ only grows as $(a/a_{eq})^2$ *outside* the Jeans scale in the radiation epoch. Once the photon fluctuations stop growing due to pressure, the CDM can only grow logarithmically. The potential thus decays essentially as $a^{-2}$ until matter domination.

Moreover equation (5) is merely that of a damped, forced oscillator with time dependent frequency and can be readily solved using the WKB approximation and the Green's function method. We discuss in detail elsewhere[4,5] the derivation of these equations and analytic solutions, including adding in the effects of anisotropic stress. On the other hand, the physical features of these solutions, including their qualitative dependence on cosmological parameters, can be obtained by merely examining equation (5). The homogeneous equation $F(\eta) = 0$ tells us that solutions are of the form $(1 + R)^{-1/4}\cos\phi$ or $(1 + R)^{-1/4}\sin\phi$ where the phase is given by

$$\phi(a, k) = k \int_0^a c_s d\eta,$$

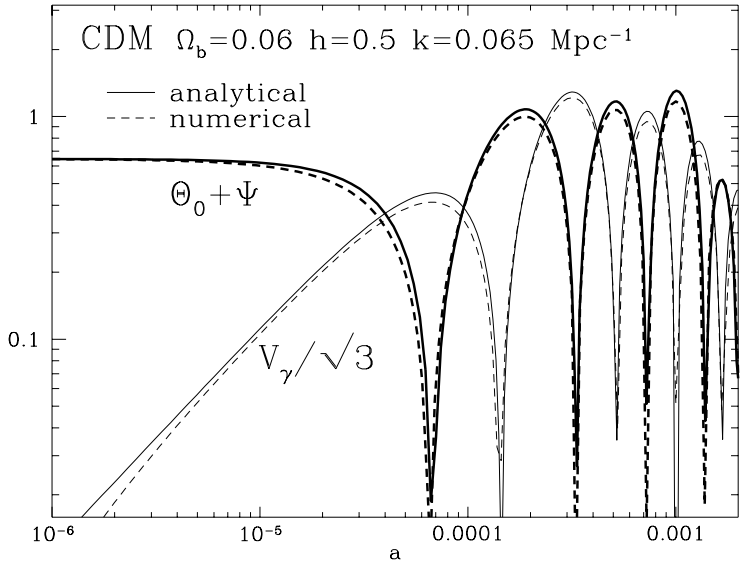

**Figure 3.** The CDM adiabatic effect. Photon pressure causes photon fluctuations to oscillate as an acoustic wave. Notice that the photon dipole is out of phase with, and becomes eventually smaller than the monopole, as we approach photon-baryon equality. The $\sqrt{3}$ is there to account for the three degrees of freedom in the dipole. This implies that fluctuations at recombination, which are directly visible as anisotropies today, will correspond to peaks in the monopole and troughs to peaks in the dipole.

with the photon-baryon sound speed

$$c_s^2 = \frac{1}{3}\frac{1}{1+R}.$$

The phase integral can in fact be performed analytically.[5] As one might expect, if the sound speed were constant, the dispersion relation for these acoustic waves is $\omega = kc_s$.

The time dependence of the gravitational driving force mimics $\cos\phi$ in the adiabatic case and $\sin\phi$ in the isocurvature case as we have seen and thus stimulates the corresponding mode of temperature perturbations. The dipole can be obtained from the photon continuity equation $V_\gamma = -3/k[\dot\Theta - \dot\Psi]$ and is of the form $(1+R)^{-3/4}\sin\phi$ and $(1+R)^{-3/4}\cos\phi$ for the adiabatic and isocurvature modes respectively. Because of the feedback of the dipole to the monopole, i.e. the oscillatory structure of the photon evolution equations, the continuity equation does *not* imply that $V_\gamma \propto \Theta_0/k$. Conventional wisdom that the monopole dominates at smaller scales than the dipole is misguided.

Notice also that the dipole is $\pi/2$ out of phase and increasingly suppressed with respect to the monopole (see Fig. 3). The magnitude of this suppression increases with $R$ and thus is greater for larger $\Omega_b$ and $h$. Of particular interest is the phase at recombination, since the temperature fluctuations at this time will be frozen in and observable today as anisotropies. Modes which were caught at the peak of the monopole will correspond to peaks in the photon power spectrum $P_\gamma$. The zeros of the monopole will be partially filled in by the dipole and represent troughs in the power spectrum. This simple analysis is enough to understand the broad structure of the Doppler peaks.

2.5. *Locations and Heights of the Doppler Peaks*

To gain deeper insight into their structure, we must examine equation (5) somewhat more carefully. In particular, one would like to know where the Doppler peaks are located, what their heights are, and how they depend on cosmological parameters.

Let us first tackle the simpler question of their location. As noted above, the $m$th peak in the adiabatic spectrum is at $\phi(a_*, k_m) = m\pi$, and in the isocurvature spectrum at $(m - 1/2)\pi$ where $m$ is an integer $\geq 1$. In Fig. 4, we plot the scale $k_p$ at which $\phi(a_*, k_p) = \pi$ for recombination at $z_* = 1000$. In a low $\Omega_b$ universe, recombination occurs well before photon-baryon equality, and the sound speed is roughly independent of $\Omega_b$. This implies that the location of the peaks will not depend strongly on $\Omega_b$ as is well known.

Further examination, shows that the condition $\phi = \pi$ requires that in the low $\Omega_b$ limit, the wavenumber at the peak $k_m \propto h$ (unless $\Omega_0 \ll 0.1$ where recombination happens well in the radiation dominated epoch in which case $k_m \propto h^0$). Today, this fluctuation is seen as anisotropy in the multipole $\ell \sim k_m r_\theta$, which corresponds to the angle subtended by the scale $k_m$ at the distance of last scattering surface, where $r_\theta = 2/H_0 \Omega_0$ in the small angle approximation. Thus, the angular location of the Doppler peaks are nearly *independent* of both $h$ and $\Omega_b$. Notice however that $r_\theta$ only equals $\eta_0$ in a flat universe. In a low $\Omega_0$ model such as PIB, the peak scale will move to higher $\ell$. This effect is just due to the more rapid deviation of geodesics in an open geometry and has the important consequence that low $\ell$ multipoles correspond to *larger* scales in open universes. These large scales are more immune to the effects of scattering and thermal history. The $\Omega_0$ dependence of the location of the peaks may also be used to "measure" $\Omega_0$ assuming the dominance of the adiabatic effect.[24]

Now let us consider the amplitude of the Doppler peaks. In order to be specific, we will now concentrate on the $\Omega_0 = 1$ CDM scenario, since in PIB the peaks are likely to be destroyed by reionization. These arguments can easily be modified to encompass PIB if desired.[5] Consider the first Doppler peak. Recall that there are essentially two, in some sense opposing effects, that contribute to the final anisotropy: adiabatic growth due to infall into potential wells and the Sachs-Wolfe effect which redshifts the photons as they climb out of these potential wells. Since adiabatic growth is determined by the balance between gravitational infall and pressure, naively one might assume that deepening the potential wells should increase the anisotropy. One way to deepen the potentials is by making equality earlier, i.e. raising $h$. However the Sachs-Wolfe effect more than counters this tendency. Therefore deepening the potential wells *reduces* the anisotropy.

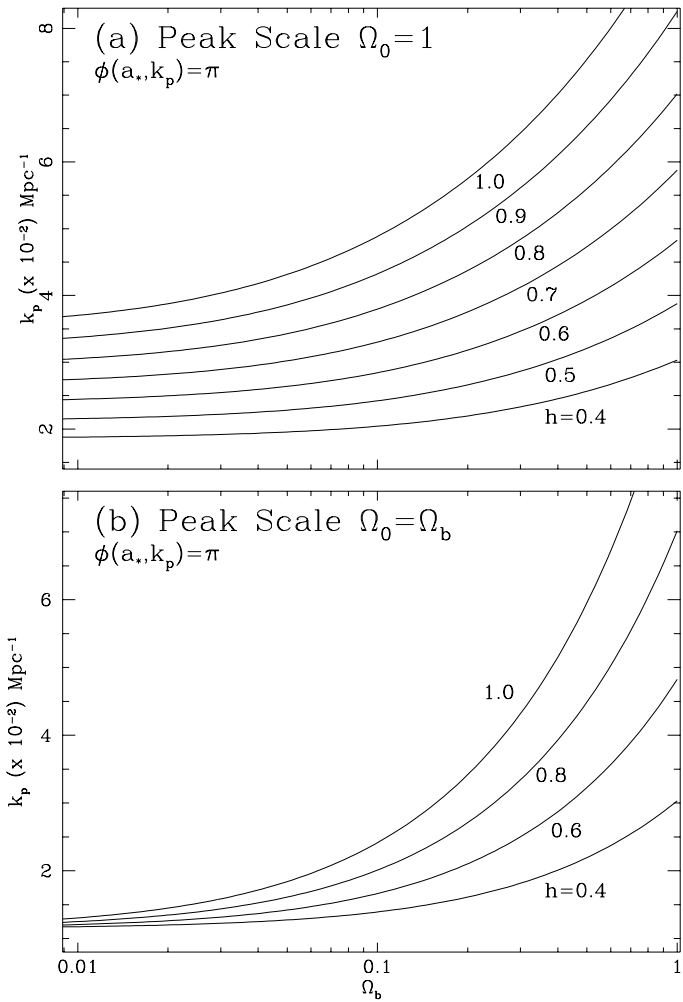

**Figure 4.** Location of the Doppler peaks. The scale of the $m$th Doppler peaks is given by the condition that $\phi = m\pi$ for adiabatic and $\phi = (m-1/2)\pi$ for isocurvature conditions. Here we plot the scale for which $\phi = \pi$ for (a) an $\Omega_0 = 1$ universe and (b) an $\Omega_0 = \Omega_b$ universe. Notice that the dependence on $\Omega_0$ and $\Omega_b$ is weak whereas it is linear in $h$.

Clearly then, the way to increase the adiabatic growth without a counterbalancing decrease through the Sachs-Wolfe effect is to lower the pressure. Since $c_s^2 = 1/3(1 + R)$ and $R \propto \Omega_b h^2$ this can be accomplished by increasing $\Omega_b h^2$. Anisotropies therefore monotonically increase with $\Omega_b$ as is well known. What about the $h$ dependence? The two tendencies discussed above oppose one another. Notice however that the pressure becomes independent of $\Omega_b h^2$ as this quantity goes to zero. Therefore at low $\Omega_b$, increasing $h$ decreases the anisotropy, whereas at high $\Omega_b$, it increases the anisotropy. The crossover point is near the nucleosynthesis value of $\Omega_b \approx 0.05$ making the $h$ dependence relatively weak for CDM.

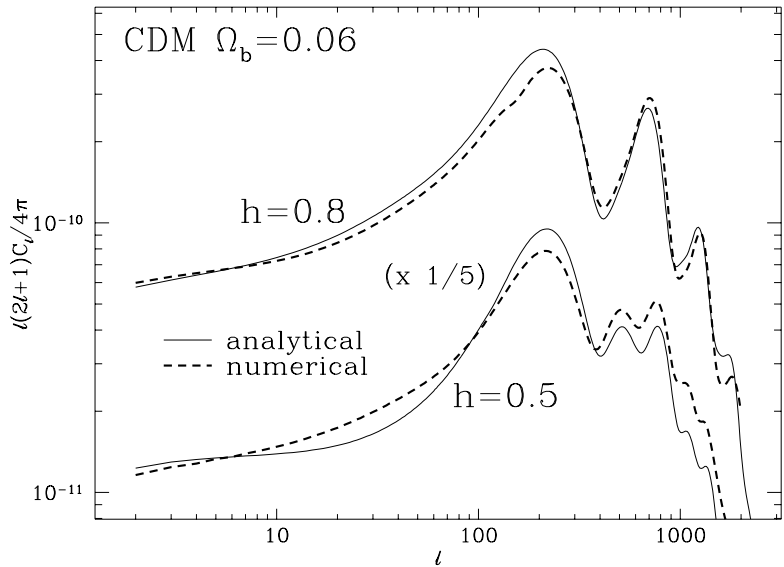

**Figure 5.** The CDM anisotropy spectrum. Notice that the analytic treatment accounts quite well for the structure of the Doppler peaks and obtains results to $\sim 10-15\%$ in temperature. The dependence on cosmological parameters is described in the text.

Now let us consider the general structure of all the peaks. Since equation (5) is merely an oscillator in a gravitational well, the (positive) compressional phase is enhanced by infall whereas the (negative) expansion phase is suppressed. The result is that the even numbered peaks are increasingly suppressed in relation to the odd ones the more important gravity is compared with pressure. For high $\Omega_b h^2$ universes, the even numbered peaks may be completely hidden. Since the dipole is also suppressed in these models, there will be a significant decrement between the first and third peaks. The full analytic solution of equation (5) reproduces all these features to 10-15% in temperature.[4] For example in Fig. 5, we compare the numerical and analytic results for the final anisotropy spectrum $(\Delta T/T)^2_{rms} = \sum (2\ell + 1) C_\ell W_\ell / 4\pi$ where $W_\ell$ is the experimental window function.

*2.6. Diffusion Damping*

Now let us return to our general discussion applicable to both PIB and CDM. Thus far, we have been assuming that tight coupling holds the photons precisely in place in the fluid. In reality, the photons random walk through the baryons. This diffusion causes photons from different temperature regions to mix and form anisotropies just as in the free streaming limit. Unlike the free streaming situation, rescattering rapidly isotropizes the distribution such that for any multipole $\ell > 1$, $\Theta_\ell \propto \exp(-\tau)$ where $\tau$ is the optical depth to Compton scattering. Therefore photon diffusion suppresses fluctuations rapidly once the perturbation comes within the diffusion length. Extending equation (5) by taking into account second order corrections to the tight coupling limit, we obtain the well known result that the above solutions must be replaced by

$$[\Theta_0 + \Psi](\eta) \to [\Theta_0 + \Psi](\eta) \exp\{-[k/k_D(\eta)]^2\}$$

where

$$k_D^{-2}(\eta) = \frac{1}{6} \int d\eta \frac{1}{\dot\tau} \frac{R^2 + 4(1+R)/5}{(1+R)^2}.$$

This analysis even works through recombination as long as we follow the ionization fraction $x_e(\eta)$ properly: $\dot\tau = x_e n_e \sigma_T a$ with $n_e$ the electron number density and $\sigma_T$ the Thomson cross section. This then accounts for the finite thickness of the last scattering surface. It is worthwhile to note that since $\tau(a)$ becomes roughly independent of the cosmological parameters $\Omega_b$ and $h$ through recombination,[25] the final diffusion length $k_D^{-1}(\eta_*) \propto h^{-1}$ increases more dramatically with $h$ at recombination than one might naively expect. This also implies that the "finite thickness" cutoff in $\ell$ is roughly independent of $h$.

Photon diffusion also has interesting consequences for the baryons. Well inside the horizon $\Delta_\gamma = 4\Theta_0$. Since tight coupling requires that $\dot\Delta_b = {}^3\!/\!_4 \dot\Delta_\gamma$, photon diffusion damping will also damp the *adiabatic* component of the baryon fluctuations. This is known as Silk damping. Since the entropy $S$ remains constant, the baryons are left with perturbations $\Delta_b = S$ after damping.

Let us put this together with the large scale behavior to form the PIB matter transfer function $P(k) = [T(k)S(k)]^2$. On scales larger than the Jeans length, the matter fluctuations have a $k^2 - 3K$ tail and grow as $a$ in the matter dominated epoch. Below this scale, the perturbations have damped oscillations around the constant $S$. Since the Jeans scale goes to a constant in the matter dominated epoch, this implies that the transfer function will have a significant peak at this maximal Jeans scale.

We can in fact deduce the properties of the oscillatory regime from our simple analysis. Since the amplitude of the oscillations is taken from its value at Jeans crossing $a_J \propto k^{-1}$, the fact that $\Delta_b \propto (1 - 3a/4a_{eq})S$ implies that the oscillation amplitude, given by the second term, will be $\propto k^{-1}$. This means that the oscillations in the transfer function decrease as $[1 + R(a_*)]^{-1/4} k^{-1}$ until they absolutely disappear for $k \gg k_D(a_*)$ leaving a constant tail $S$. This constant tail however can grow after Compton drag becomes unimportant and is thus sensitive to the ionization history (see §2.7). Thus the amplitude of fluctuations at galaxy scales and prominence of the peak can be tuned through these "environmental" parameters.

Contrast this with the CDM transfer function. The baryons fall into CDM wells after Compton drag becomes negligible. Outside the Jeans scale in the radiation dominated limit, CDM perturbations grow with the photons as $\Delta_{CDM} \propto (a/a_{eq})^2$ until Jeans scale crossing. Thus the transfer function is flat at large scales and goes to $k^{-2}$ for modes that cross the Jeans scale in the radiation dominated limit.

Since large scale structure measurements indicate $n \approx -1$ at intermediate scales $10^{-2} \lesssim k/h \lesssim 1$ Mpc$^{-1}$, which fall just below the maximal Jeans scale, PIB must have an *initial* spectrum of $n \approx -1$ to satisfy observations. CDM on the other hand, due to the lack of entropy fluctuations, can have Harrison-Zel'dovich initial conditions of $n \approx 1$ and still satisfy the constraints. More worrisome is the fact that large scale structure measurements fall in the oscillatory regime of the PIB transfer function. Indications of a smooth power spectrum by Peacock & Dodds[12] may make it difficult for many PIB models. Note however that reionization can help the situation as we shall see in the next section since the oscillations are damped as $[1 + R(a_*)]^{-1/4}$, the diffusion length grows, and the maximal Jeans scale reaches its full matter dominated value.

### 2.7. The Doppler Effect

As shown above, the PIB model requires a spectrum with quite a large amount of *small* scale power $P(k) \propto k^{-1}$ in contrast with the CDM model $P(k) \propto k^{n-4} \approx k^{-3}$ to form large scale structure. In this model, a first generation of objects is likely to form right after standard recombination which could reionize the universe.[6] In fact, the large amount of small scale power also entails adiabatic oscillations far too large for present observations on the arcminute to degree scale.[26] Reionization is not only natural but also necessary for this model since scattering damps anisotropies by $\exp(-\tau)$ as we have seen. These considerations are of course also applicable to reionized CDM models as well.

If the universe is reionized, the photon diffusion length continues to grow and we must consider how temperature perturbations are generated inside the diffusion length. Although intrinsic adiabatic fluctuations in the photons have damped away, perturbations can be regenerated through the Doppler effect. In this case the photon velocity (dipole) is no longer equal to the baryon velocity and scattering thus leads to a Doppler shift. The baryon velocity can also grow after Compton drag has become negligible: $z_d \approx 160(\Omega_0 h^2/x_e^2)^{1/5}$, which is generally earlier than last scattering $z_* \approx 30(\Omega_0/0.1)^{1/3}(0.05/x_e\Omega_b h)^{2/3}$ in reionized scenarios. Note therefore that the amplitude of the final matter fluctuations and its dependence on drag and cosmological parameters can be described as $D(\eta_0)/D(\eta_d)$, where $D$ is the growth factor in an open universe.[27] This is useful in estimating $\sigma_8$, the mass overdensity on the $8h^{-1}$ Mpc scale. In Fig. 6, we show the dependence of the matter transfer function on ionization $x_e$.

As photons random walk through the electrons, they pick up Doppler shifts from Compton scattering. These fluctuations are canceled out as the photons continue to random walk through many wavelengths of the baryon perturbation. Statistically only the last scattering event is important. However, cancellation occurs even through the last scattering surface since it is of finite thickness.

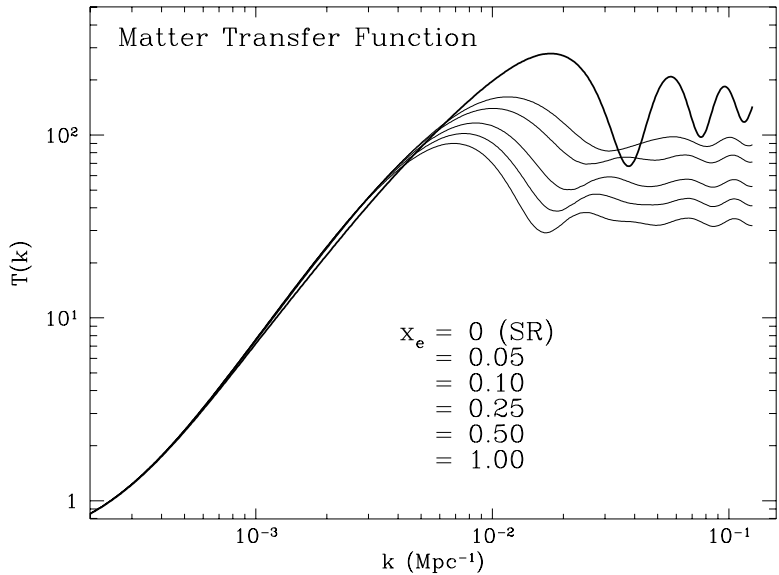

**Figure 6.** The PIB matter transfer function. The prominent peak in the transfer function is at the Jeans scale at last scattering. Later last scattering implies that the Jeans scale can grow to its full matter dominated value. The constant tail in the transfer function grows after the matter is released from Compton drag and thus is larger for low ionization models. The adiabatic oscillations present in the standard recombination model (SR) are damped out in the reionized models.

This severe cancellation of the Doppler effect at scales smaller than the thickness of the last scattering surface makes ordinarily negligible effects appear: second order contributions from mode coupling between densities and velocities (the Vishniac effect[28]) and the (small scale) ISW effect.[29] All these contributions can be calculated analytically by employing thick last scattering techniques which rely on severe statistical cancellation[29,30,31] (see Fig. 7). The basic idea is that only modes with fluctuations perpendicular to the line of sight will survive. Parallel to the line of sight, crests and troughs of the perturbations will cancel. A more detailed discussion of the cancellation mechanism can be found elsewhere.[31] Moreover, no recombination occurs in these scenarios so that last scattering happens gradually due to the decrease in electron density from the expansion. Thus the last scattering surface will be comparable in thickness to the horizon at last scattering which nearly closes the window for the adiabatic effect.

Aside from the Vishniac effect, which can often be avoided due to its quadratic dependence on normalization, these considerations imply that anisotropies will be increasingly suppressed as the last scattering surface becomes thicker. The smallest fluctuations that can be obtained are for the latest last scattering, i.e. the fully ionized $x_e = 1$ model. Unfortunately, assuming a reasonable thermal history for the electron temperature $T_e$, this model is ruled out by constraints on spectral distortions in the CMB. Ionization implies a hot intergalactic medium with $T_e \gtrsim 5000$K. By Compton scattering off these hot electrons, the CMB photons are boosted in frequency leading to a characteristic Raleigh-Jeans suppression and Wein tail enhancement described by the Compton-$y$ parameter constrained by COBE FIRAS[9]

$$y \approx 8.4 \times 10^{-7} \tau \left(\frac{T_e}{5000\text{K}}\right)\left(1 - \frac{3}{5}\frac{T_0}{T_e}z_i\right) < 2.5 \times 10^{-5},$$

if the universe has been ionized to a fraction $x_e$ since the ionization redshift $z_i$. To get the minimum anisotropies one must tune the ionization to be as high as possible without violating the COBE FIRAS constraint.

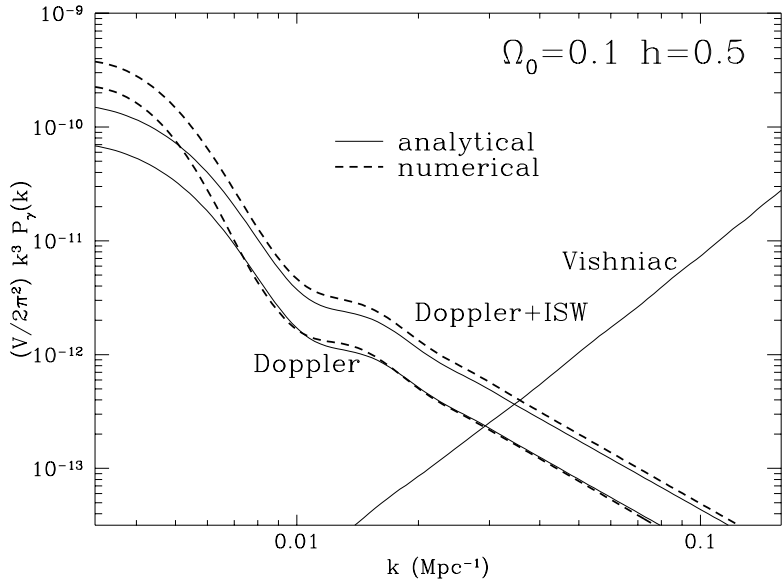

**Figure 7.** The PIB small scale temperature power spectrum $P_\gamma = |\Theta+\Psi|^2_{rms}$. Analytic calculations involving severe cancellation are valid at scales small compared with the thickness of the last scattering surface. Notice that whereas the Doppler and ISW effect have the same scale dependence, the Vishniac effect becomes dominant at extremely small scales.

## 3. The Nurture of Anisotropies

*3.1. Environmental Factors: Thermal History*

To what extent then can the natural difference between the PIB and CDM scenarios be used to distinguish the two. Can flexible "environmental factors" such as the thermal history since $z = 1000$ assimilate the two? In §2, we have seen that despite formal similarities in the mechanisms of anisotropy and matter fluctuation generation, the PIB model generally *does* make quite different predictions from CDM in its simplest form. In particular, large scale structure formation tells us that the PIB model must have $n \approx -1$ *initially* on such scales *and below*, assuming pure power law initial conditions, which implies that reionization is both natural and necessary. However, large scale structure measurements do not have enough dynamic range yet to clearly distinguish between an $n = -1$ PIB model and a processed $n = 1$ CDM model, although this issue may be settled soon.[12] What about the amplitude of matter fluctuations described by $\sigma_8$? Unlike the CDM model, the final amplitude of the transfer function is dependent on ionization history, for there are no CDM wells into which the baryons may fall. This implies that one can tune the ionization history to match observations in this respect. Models with $n \gtrsim -0.5$ are in fact ruled out because the necessary ionization history violates COBE FIRAS constraints on spectral distortions. However, for the preferred value of $n \approx -1$ for large scale structure formation, such an ionization history is not yet in conflict with spectral distortions. Further complications arise due to the possibility that early structure formation dumped most of the baryons in black holes which behave effectively as CDM. The general picture however remains the same. We refer the interested reader to our more detailed treatment.[8] To summarize the status of matter fluctuations: PIB can be made to look almost indistinguishable from CDM by a judicious selection of the ionization history.

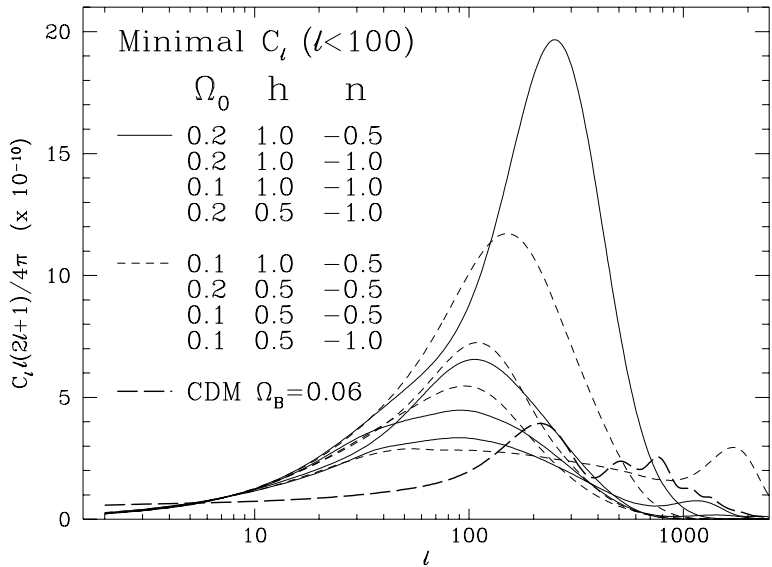

**Figure 8.** Minimal anisotropies in the PIB model consistent with COBE FIRAS and Vishniac constraints. Notice that every model has significantly larger anisotropies between $\ell$ of $10 - 100$ than CDM. All PIB models have similar slopes at the COBE DMR scale $n_{\text{eff}} \approx 2$.

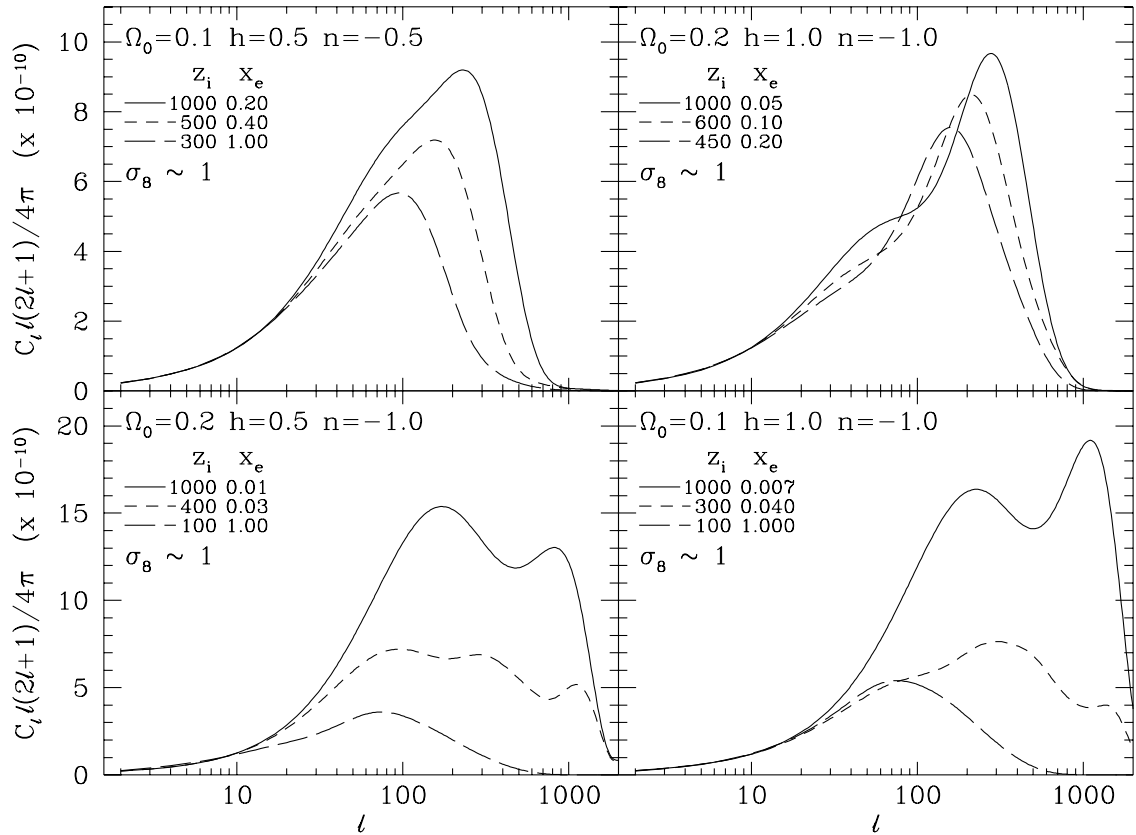

**Figure 9.** PIB models satisfying $\sigma_8 \approx 1$. Aside from low optical depth models which predict enormous anisotropies compared with CDM, all PIB models have only a single feature at the thickness scale of the last scattering surface. Thus the amplitude and structure are quite distinct from the CDM spectrum.

Obviously then, hope for distinguishing PIB lies mainly in combining CMB predictions with large scale structure requirements for the initial spectral index and ionization history. In fact there are several clear distinctions between the PIB and CDM anisotropy predictions. The large amounts of small scale power make degree scale anisotropies significantly larger in the PIB model even for the minimal (and excluded) fully ionized model. In Fig. 8, we show the minimal CMB anisotropies for models consistent with COBE FIRAS limits and constraints on the Vishniac effect.[8] Note that to be conservative, in choosing these "minimal" models we have for the time being relaxed large scale structure constraints, i.e. $\sigma_8 \neq 1$ and $-1 \lesssim n \lesssim -0.5$. Complications such as the dependence of the adiabatic effect on ionization history and drag do not change this picture significantly.[8]

The requirement that the model be reionized also tells us that the PIB anisotropy spectrum is dominated by the Doppler effect which only has a *single* feature at the thickness scale of the last scattering surface. Contrast this with the adiabatic effect, dominant in the standard recombination CDM model, which produces a spectrum of peaks. As might be expected, the exception to this rule is a PIB model with low optical depth (early last scattering) where the adiabatic effect can create additional features.[8]

Finally, large scale anisotropies also provide a reasonably robust test of PIB. An open geometry implies that a given scale corresponds to a *larger* $\ell$ than in a flat model. This implies that large scale anisotropies $\ell \lesssim 30$ are an excellent discriminator between PIB and CDM since they are independent of uncertainties such as thermal history. Large scale anisotropies in the PIB model are significantly steeper in slope than the CDM model and corresponds to $n_{\text{eff}} \approx 2$ at the COBE DMR scale for all PIB models[32] as explained in §3.2. However, large scale anisotropies may suffer more from uncertainties in initial conditions as we shall see in §3.2.

Since the observational status of anisotropies is rapidly changing, we will not try to make a definitive comparison with experiments here. However it is worthwhile to note that if the low fluctuations reported by SP91[33] are confirmed, *all* PIB models are ruled out,[8] basically since even the fully ionized model is inconsistent,[34] for models with initial slope $n$ fixed by large scale structure. PIB may also run into conflict with recent findings that $n_{\text{eff}} \approx 1$ at the COBE DMR scale.[35] On the other hand all PIB models fare favorably with respect to the high detections by the Tenerife[36] and MAX experiments.[37] For more details on comparisons to these and other experiments, see our more complete treatment.[8] Finally let us present some more "realistic" models that are consistent with large scale structure constraints of $\sigma_8 \approx 1$. In Fig. 9, we plot the numerical results for CMB anisotropies where the ionization history has been fine tuned to match this constraint. Again, we see that all models predict significantly larger degree scale anisotropies than CDM.

*3.2. Eugenics: Designer Initial Conditions*

The single most troublesome caveat for the findings of the last section is that there are in reality *two* free functions in the PIB model. In addition to uncertainties in the ionization history, there is no mechanism of fluctuation generation to fix the initial conditions. As discussed in §2.1, this problem is compounded by the ambiguities of an open universe. The curvature of the universe introduces a natural scale to the problem, and thus a pure power law may not be realistic. Obviously if we take the extremely pessimistic stance and allow completely *arbitrary* initial conditions, the model loses most of its predictive power. In this view, large scale structure only fixes $n \approx -1$ at galaxy to cluster scales and bears no *a priori* relation to the COBE DMR fluctuations. Even in this case, we can require that degree scale CMB anisotropies be consistent with large scale measurements which are on comparable scales. The verdict for this sort of test will have to await improvements in degree scale anisotropy measurements. However it is possible that a definitive statement can be made even on these grounds if anisotropies turn out to be low since this would also imply a low $\sigma_8$. Of course no comparison can be made if the bias is not only large but also an arbitrary function of scale. This though seems like a remote possibility in the PIB model.[38]

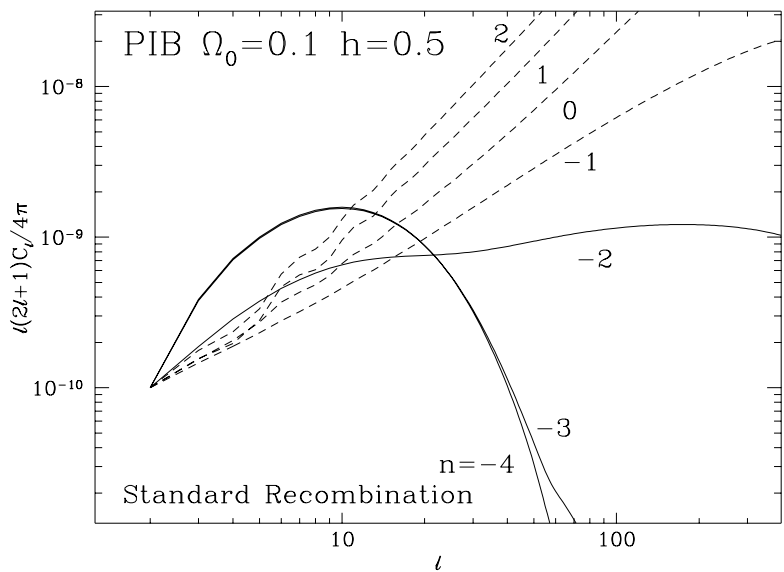

**Figure 10.** Dependence of PIB anisotropies on the initial power law. For red spectra $n < -3$, the dominance of curvature scale perturbations makes the final anisotropy independent of the initial spectrum. Similarly, for very blue spectra $n \gg -1$, the dominance of the smallest sized perturbations make anisotropies again only weakly dependent on $n$, asymptotically approaching constant $C_\ell$. The critical index of $n \approx -1$ for structure formation lies in the slowly varying regime where $C_\ell \propto \ell^{-1}$ or $n_{\text{eff}} \approx 2$. This also implies that PIB spectra with $n \approx 1$ are not terribly sensitive to ambiguities at the curvature scale.

The designer spectrum scenario described above cannot really even be properly called a model as it has almost no predictive power. A more reasonable alternative is a power spectrum which only behaves peculiarly on or above the curvature scale. To build some intuition about the possible effects of altering the spectrum at the curvature scale, let us continue to assume a pure power law but see how anisotropies change with spectral index $n$ (see Fig. 10). Consider first extremely red spectra (in fact divergent in the absence of a cutoff) with $n \lesssim -3$. Even these spectra have an $\ell$ space cutoff at very low $\ell$. Moreover, the $\ell$ dependence of the anisotropies does not depend on $n$ in this regime! This corresponds to the fact that there are no fluctuations above the curvature scale in a random field model (see §2.2). No matter how such a red power spectrum is constructed, the result is always the same: all fluctuations lie at the curvature scale and thus anisotropies mainly appear at the angle subtended by the curvature scale at last scattering.

Notice also that at the opposite extreme, blue spectra $n > -1$ seem to slowly converge to $\ell(2\ell+1)C_\ell \propto \ell^2$, with $\ell(2\ell+1)C_\ell \propto \ell$ ($n_{\text{eff}} \approx 2$) approximately in the interesting regime of $-1 < n < 0$. As $n$ increases, eventually anisotropies will be dominated by the smallest scale where fluctuations still exist, e.g. the damping scale $k_D$ for standard recombination. This invalidates the angle to distance relation $kr_\theta \sim \ell$ employed in §2.5 and thus alters the $k$ to $\ell$ mapping, i.e. power from high $k$ modes bleed into the low $\ell$ multipoles leading to a less steep $C_\ell$. It is easy to show that this implies $C_\ell$ is a constant in $\ell$ for $n \gg 1$ as required.[5] This is not a peculiarity of the open geometry or isocurvature conditions. It does not appear in standard CDM since an adiabatic $n=1$ spectrum corresponds to an isocurvature $n=-3$ spectrum below the curvature scale. This also explains why even though a Harrison-Zel'dovich CDM model, equivalent to an isocurvature $n=-3$, has $n_{\text{eff}}=1$, a PIB model with $n=-1$ does *not* have $n_{\text{eff}}=3$ but rather $n_{\text{eff}}=2$.

The upshot is that for the interesting value of $n \approx -1$ anisotropies are not very sensitive to ambiguities in defining fluctuations at the curvature scale. Only for the extreme situation where the power spectrum breaks toward near divergence at $\tilde{k} \lesssim \sqrt{-K}$ will curvature scale contributions matter. Moreover if the spectrum really has most its power at scales $\tilde{k} < \sqrt{-K}$, the small scale power would be insufficient to form large scale structure if normalized to the COBE DMR detection. Only by fine tuning the initial conditions to have exactly the right amount of large scale power can we significantly alter the conclusions for the COBE DMR slope and degree scale anisotropies.

## 4. Discussion

Hopefully in the course of this rather informal and intuitive discussion, we have dispelled some common myths related to CMB anisotropies. (The definition of "common myth" is, of course, that we ourselves believed them at some time in the past!) It is often assumed that the structure of microwave background anisotropies is too complicated to lend itself to simple analytic understanding. Moreover, multi-fluid gauge invariant perturbation theory[39] has been viewed as a formalism that is difficult to interpret physically. On the contrary, we show here that it is easy to interpret the evolution of perturbations, including CMB anisotropies, in the formalism. Moreover minor extensions to this analytic treatment can account *quantitatively* for all these effects at the 10-15% level.[4,5] The simplicity of this approach is mainly due to the existence of a Newtonian potential $\Psi$, the freedom to choose gauge invariant variables that simplify the physics,[5] and our desire not to obscure the physics with complicated mathematical analysis!

More concretely, we have shown how these analytic arguments make clear the often forgotten fact that the Doppler peaks are due to adiabatic oscillations in the photon-baryon fluid and *not* due to a Doppler shift from electrons at the last scattering surface. They also serve to bring out the physical mechanism behind their dependence on cosmological parameters.[40] The true Doppler effect is only important below the diffusion (or thickness) scale where the tight coupling breaks down. These principles allow one to state why a reionized model such as the standard PIB scenario does not exhibit the same oscillatory structure. Here, the diffusion length has grown to the extent that the Doppler rather than the adiabatic effect dominates degree scale anisotropies.

Furthermore, since diffusion does not Silk damp entropy perturbations, the PIB model has far more power on small scales than the CDM model. Moreover because of the mechanism that generates large scale density fluctuations for isocurvature initial conditions, a power index at galaxy scales $n$ ($\approx -1$) would correspond to a large scale index of $n + 4$ ($\approx 3$) if compared with an adiabatic model. This implies that a PIB model which has a small scale slope designed to match large scale structure will have anisotropies increasing with $\ell$ much more rapidly than a similarly constructed CDM model. Even in the minimal fully ionized model, these large fluctuations may already be in conflict with degree scale measurements and the COBE DMR slope. The only way to escape these conclusions is to assume a very special form for the initial power spectrum. Not even ambiguities as to how to define the power spectrum above the curvature scale seem able to modify these results significantly without fine tuning and drastic assumptions.

## 5. Acknowledgements

I would like to thank my collaborator on these efforts, N. Sugiyama. The ideas presented here are in fact taken from our previous[8,29] and forthcoming[4,5] work. I would also like to acknowledge useful conversations with E. Bunn, K. Gorski, M. Kamionkowski, D. Scott, J. Silk, R. Stompor, and M. White. This work was partially supported by an NSF Fellowship.